\documentclass[twocolumn,prb,citeautoscript,superscriptaddress,8pt]{revtex4-2}

\usepackage{graphicx}
\usepackage{gensymb}
\usepackage{color}
\usepackage[dvipsnames]{xcolor}
\usepackage{float}
\usepackage{upgreek}
\usepackage{bm}
\usepackage{amsmath,amssymb}

\begin{document}

\title{Towards high spatial resolution magnetic imaging with a compact practical quantum diamond microscope}

\author{Kevin J. Rietwyk}
\email{kevin.rietwyk@rmit.edu.au}
\affiliation{School of Science, RMIT University, Melbourne, VIC 3001, Australia}

\author{Alex Shaji} 
\affiliation{School of Science, RMIT University, Melbourne, VIC 3001, Australia}

\author{Islay O. Robertson}
\affiliation{School of Science, RMIT University, Melbourne, VIC 3001, Australia}

\author{Alexander J. Healey}
\affiliation{School of Science, RMIT University, Melbourne, VIC 3001, Australia}

\author{Priya Singh}
\affiliation{School of Science, RMIT University, Melbourne, VIC 3001, Australia}

\author{Sam C. Scholten}
\affiliation{School of Science, RMIT University, Melbourne, VIC 3001, Australia}

\author{Philipp Reineck}
\affiliation{School of Science, RMIT University, Melbourne, VIC 3001, Australia}
\affiliation{ARC Centre of Excellence for Nanoscale BioPhotonics, School of Science, RMIT University, Melbourne, VIC 3001, Australia}

\author{David Broadway}
\email{david.broadway@rmit.edu.au}
\affiliation{School of Science, RMIT University, Melbourne, VIC 3001, Australia}

\author{Jean-Philippe Tetienne}
\email{jean-philippe.tetienne@rmit.edu.au}
\affiliation{School of Science, RMIT University, Melbourne, VIC 3001, Australia}

\begin{abstract} 
Widefield quantum diamond microscopy is a powerful technique for imaging magnetic fields with high sensitivity and spatial resolution. However, current methods to approach the ultimate spatial resolution ($<500$\,nm) are impractical for routine use as they require time-consuming fabrication or transfer techniques to precisely interface the diamond sensor with the sample to be imaged. To address this challenge, we have designed a co-axial sensor holder that enables simple, repeatable sensor-sample interfacing while being compatible with high numerical aperture (NA) optics. With our new design we demonstrate low standoffs $<500$\,nm with a millimetre sized sensor. We also explore the relationship between spatial resolution and NA spanning from 0.13 to 1.3. The spatial resolution shows good agreement with the optical diffraction limit at low NA but deviates at high NA, which is shown to be due to optical aberrations. Future improvements to our design are discussed, which should enable magnetic imaging with $<500$\,nm resolution in an accessible, easy-to-use instrument.

\end{abstract}

\maketitle 
The widefield quantum diamond microscope (QDM) is an emerging, versatile tool for imaging the magnetic field from materials and devices~\cite{Levine2019,Scholten2021}. It is based on the nitrogen-vacancy (NV) centre in diamond, a fluorescent point defect with a well-known electronic structure that is sensitive to the magnetic field and can be probed using optically detected magnetic resonance (ODMR) spectroscopy~\cite{Doherty2013,Rondin2014}. In the QDM, a diamond with a near-surface layer of NV centres is placed close to the sample or device under study and the ODMR spectrum of the NV layer collected on a camera to provide a highly sensitive, quantitative, and calibration-free magnetic field image of the sample~\cite{Levine2019,Scholten2021}. The QDM has been employed to study magnetism in a myriad of solid-state~\cite{Toraille2018,Broadway2020,Meirzada2021,Mclaughlin2021,Chen2022,Lamichhane2023,Nishimura2023,Bhattacharyya2024}, biological~\cite{LeSage2013,Glenn2015,Fescenko2019,McCoey2020,Chen2023} and geological~\cite{Glenn2017,Farchi2017,Fu2020} samples and to characterise current distributions in electronic devices~\cite{Nowodzinski2015,Turner2020,Ku2020,Scholten2022,Kehayias2023,Garsi2024,Wen2024}. 

In the QDM, the spatial resolution is determined by the convolution of two effects~\cite{Scholten2021}. First, the standoff distance between the diamond sensor and the sample under study limits the smallest feature size resolvable in the magnetic field experienced by the NV layer~\cite{Casola2018}. Second, since the magnetic field at the NV layer is read out optically, the spatial resolution is further limited by the optical resolution of the microscope, characterised by its point spread function (PSF)~\cite{Scholten2022b,Nishimura2024}. Improving the combined spatial resolution is critical to many applications of the QDM, not only to resolve smaller spatial features but also to enhance measurement precision and accuracy (as the measured magnetic field amplitude scales with spatial resolution) and reduce associated imaging artefacts~\cite{Scholten2022b}. The simplest approach to achieve low standoffs is to deposit/fabricate the sample/device directly onto the diamond surface, with demonstrated spatial resolutions down to $\approx400$\,nm corresponding to the optical diffraction limit with a high numerical aperture objective (NA\,$\gtrsim\,0.8$)~\cite{LeSage2013,Fescenko2019,Mclaughlin2021,Lamichhane2023}. This approach works well for the study of nanoparticles or samples that can be deposited in-house~\cite{Broadway2020,Mclaughlin2021,Chen2022,Lamichhane2023,LeSage2013,Glenn2015,Fescenko2019} but for many applications it is impractical. A more flexible approach is to manually position the diamond on the sample, however, achieving an ideal standoff poses its own challenges. With millimetre-sized diamonds, imperfect flatness and contamination at the interface often result in relatively large standoffs of the order of 10\,µm~\cite{Toraille2018,Turner2020,Fu2020,Meirzada2021}. Reducing the lateral size of the diamond to tens of microns has proven effective to achieve $\lesssim$\,1\,µm standoffs~\cite{Schlussel2018,Ghiasi2023,Asif2024}, but unfortunately limits the field of view. Moreover, transferring and positioning a diamond sensor on a sample is a time-consuming process that hinders routine use and high-throughput studies. 

An alternate approach developed by Abrahams et al.~\cite{Abrahams2021} is to attach the diamond sensor to an arm that can be finely positioned and tilted relative to the sample. Using an alignment procedure based on optical interference fringes~\cite{Ernst2019}, a standoff of $\lesssim2$\,µm across a millimetre-scale diamond was repeatably achieved in standard lab conditions~\cite{Abrahams2021}. Importantly, this approach allows rapid sample exchange, with no special sample fabrication or diamond handling technique required. However, the arm design of Abrahams et al.\ is relatively bulky and prevents the use of high NA optics, thereby limiting the achievable spatial resolution. In this work, we present an improved QDM design that is compact, practical, and fully compatible with high NA objectives including oil-immersion objectives. Our QDM relies on a sensor holder employing a sleeve design that fits around the objective and makes the alignment stage co-axial with the objective, providing higher stability and better control with a small footprint. We characterise the optical resolution of the system for various objectives with NA spanning from 0.13 to 1.3. Good agreement with the diffraction limit is found for NA up to 0.8, leading to a resolution down to $\approx500$\,nm, but the resolution saturates at higher NA due to aberrations caused by imaging through the diamond. We then demonstrate how standoffs of $\approx500$\,nm can be readily achieved with our design despite its simplicity. Finally, we develop a model of aberrations to predict the maximum acceptable diamond thickness to reach diffraction-limited performance, and discuss future improvements to our design. Our work is a significant step towards a compact, practical QDM capable of magnetic imaging with a spatial resolution approaching the ultimate limit ($<300$\,nm with NA\,$=1.3$).

\begin{figure}[tb!]
    \centering
    \includegraphics[width=8.5cm]{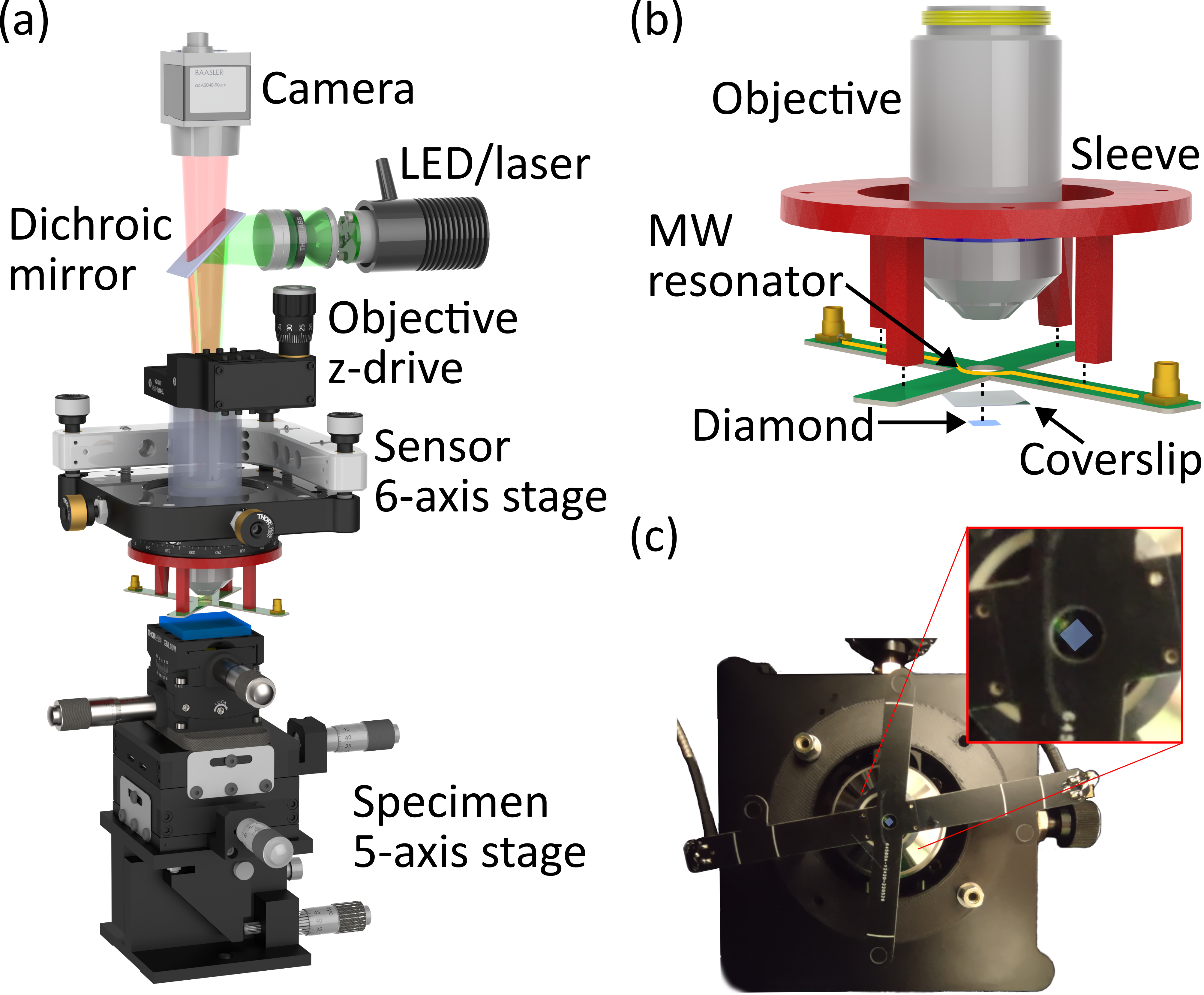}
    \caption{\textbf{A QDM with sensor sleeve for high spatial resolution.} (a) Schematic of the widefield microscope including our new sensor holder mounted on an independent 6-axis stage. Green light from an LED or laser is focused onto the diamond sensor to excite the NV layer in the diamond using an objective via a dichroic reflector. Photoluminescence from the diamond sensor is imaged using a camera. A MW field is generated using a MW resonator within the PCB and external magnetic field is applied using a positioned permanent magnet (not shown). A 5-axis stage is used to manipulate the specimen with respect to the sensor. (b) Schematic of our 3D-printed co-axial sensor holder. The diamond sensor is glued to a coverslip, fixed to a PCB with an integrated MW resonator and connectors which is attached to the sensor holder. Space between the PCB and objective permit the sensor to be translated and tilted. (c) Photograph of the sensor holder, PCB (3.2 mm aperture) and diamond sensor attached to the 6 axis-sensor stage, from below (false color was added to the diamond for emphasis).}
    \label{fig:QDM_schem}
\end{figure}

A schematic of the overall QDM is shown in Fig.~\ref{fig:QDM_schem}(a) and will be briefly described here before detailing our new sensor holder. Light from either a laser or LED is focused onto the diamond sensor to excite the NV layer using an objective on a z-drive via dichroic reflector. The diamond sensor is glued to a coverslip (using low fluorescence glue) which is fixed to a printed circuit board (PCB) with an microwave (MW) resonator around an aperture to allow optical excitation and imaging. The PCB is glued to the sensor holder which is mounted on a 6-axis stage. A MW field is generated using a MW resonator and external magnetic field is applied using an intentionally positioned permanent magnet (not shown). Several PCBs has been created with various aperture and resonator sizes to provide a compromise between a suitable viewing area for the objective and ensuring a sufficient MW flux. A PCB with a 3.2\,mm aperture was used for air objectives 4x (NA = 0.13), 10x (0.3), 20x (0.45), 40x (0.6) and 50x (0.8), shown in Fig.~\ref{fig:QDM_schem}(b). However, to position the 40x oil objective (NA = 1.3) close enough to the cover slip, a PCB with a 12 mm diameter aperture was required, see Fig. S2(b). The specimen is brought into close proximity, parallel to the diamond surface using a 5-axis stage (3 positional and 2 rotational axes). 

The sensor holder is mounted on a 6-axis stage including 3 positional and 3 rotational axes to allow for independent lateral positioning and tilting of the diamond sensor (Fig.~\ref{fig:QDM_schem}(c)). The stage (Thorlabs K6X2) allows \textpm 2\,mm lateral and \textpm 3.2\,mm vertical movement and \textpm 2\textdegree\,pitch/yaw adjustment with microradian angular control, sufficient to achieve $<100$\,nm height variation across 1\,mm. The sensor holder is attached to the stage via screws and can be easily removed and replaced to change objectives within minutes. We note that a simplified version of the sensor holder design was reported in Ref.~\cite{Shaji2024} which was only suited to low NA objectives and with limited control over the sensor position and tilt. 

\begin{figure*}[tb!]
    \centering
    \includegraphics[width=17cm]{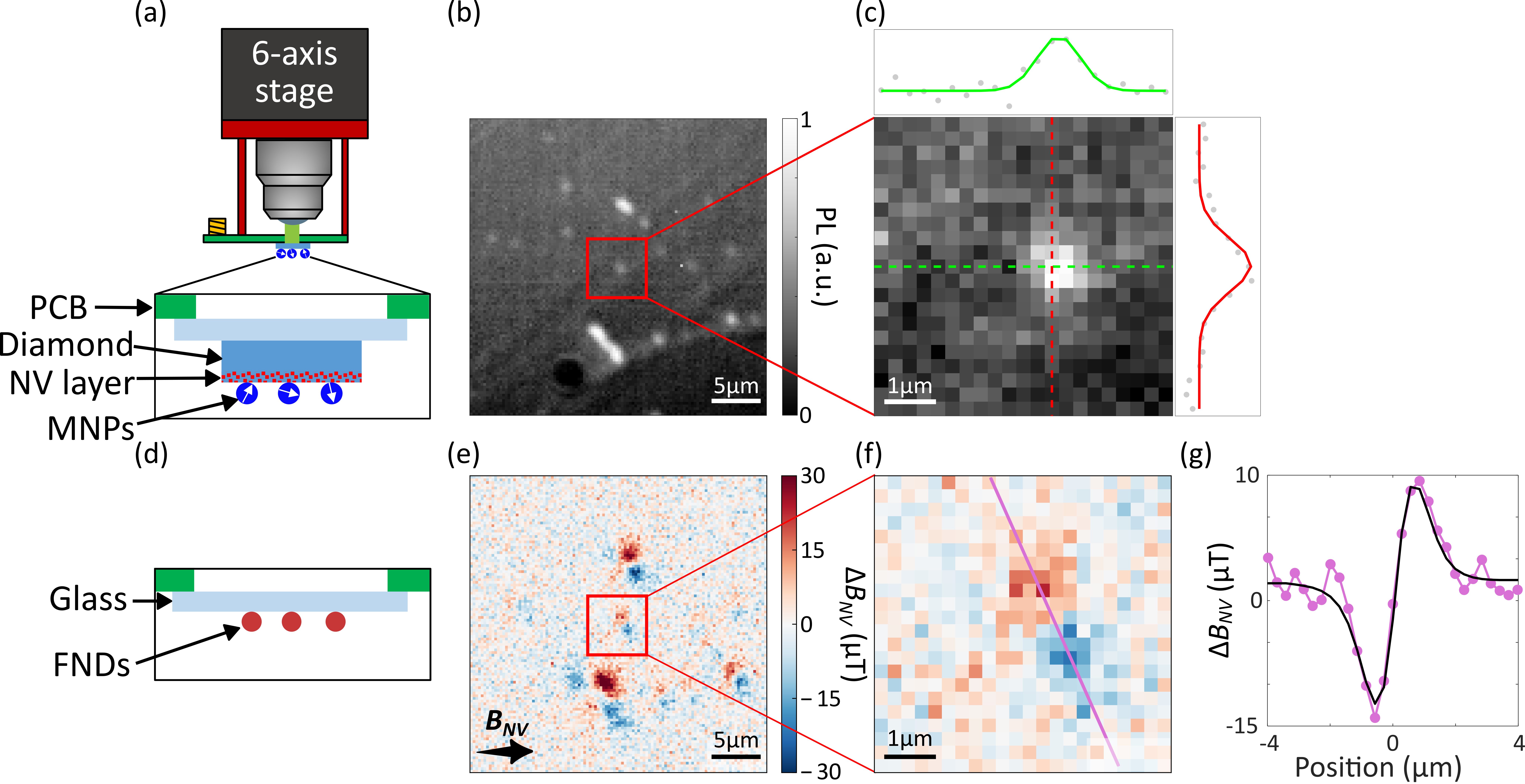}
    \caption{\textbf{Determination of the optical and magnetic resolution.} (a) Schematic of Fe\textsubscript{2}O\textsubscript{3} MNPs (\textless 50\,nm) deposited directly onto the diamond sensor, glue to a PCB and mounted on our sensor holder. (b) PL image of the diamond sensor with the Fe\textsubscript{2}O\textsubscript{3} MNPs appearing as bright features. (c) Zoom-in of a nanoparticle, shown in (b), fitted with a 2D Gaussian function to ascertain the PSF. Horizontal and vertical line profiles across the nanoparticle and line profiles of the fit (solid lines) are shown. (d) Schematic of FNDs deposited onto a glass coverslip, used for the spatial resolution analysis in Fig.~\ref{fig:FWHM_vs_NA}. (e) Magnetic field image – the stray field projected along one of the NV axes in the diamond sensor ($\Delta B_{\rm NV}$) taken at the same position as (b). (f) Zoom-in of the magnetic image in (e) for the same region as (c). (g) Line profile across the main features in the magnetic image in (f) and corresponding fit to determine the Gaussian broadening of the magnetic image based on simulations.}
    \label{fig:Det_res}
\end{figure*}

\begin{figure}[tb!]
    \includegraphics[width=8.5cm]{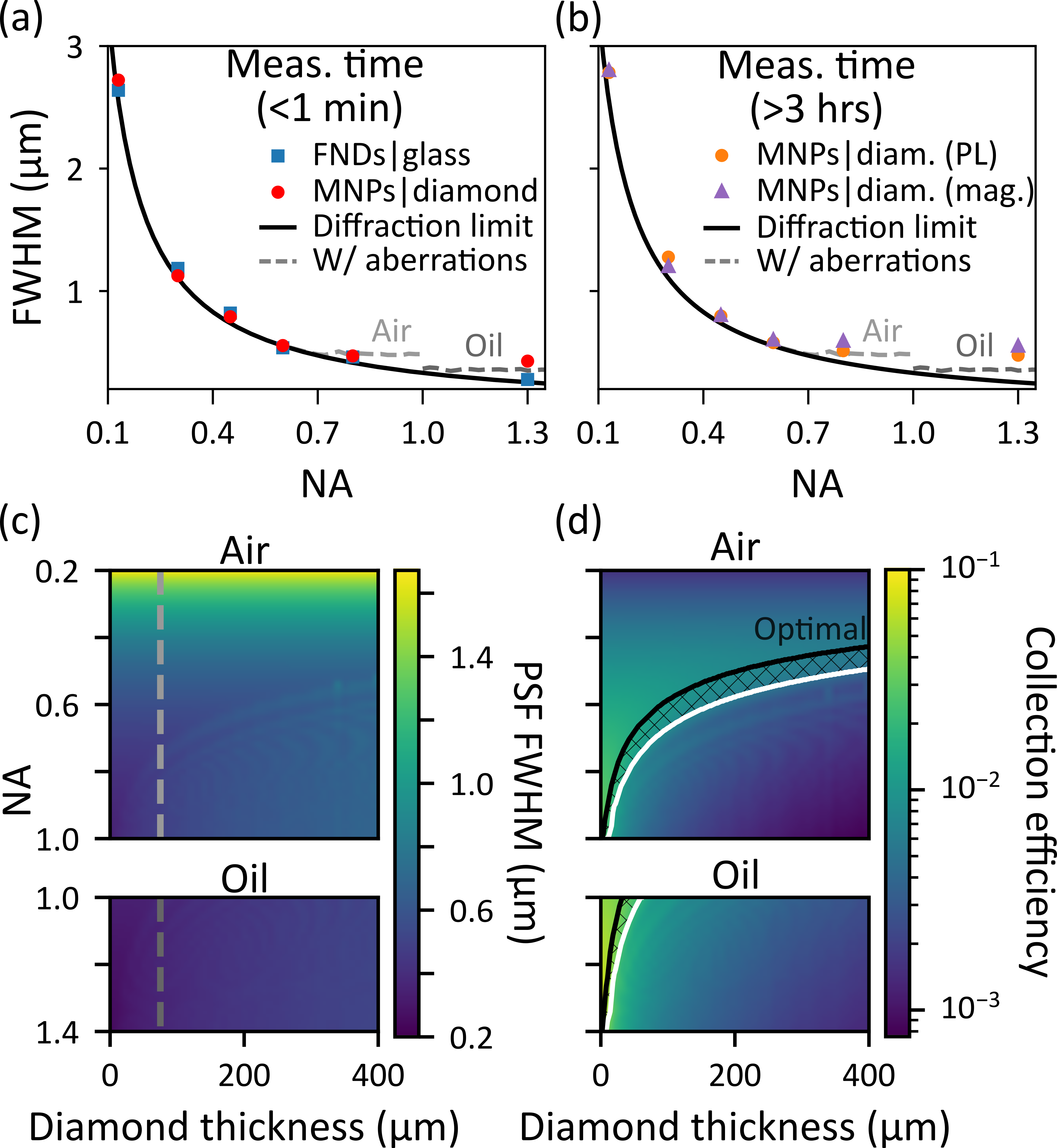}
    \caption{\textbf{Spatial resolution analysis.}
    (a) FWHM of the PSF for FNDs on a coverslip (blue) and Fe\textsubscript{2}O\textsubscript{3} MNPs deposited directly onto diamond for each objective (red), determined from quick PL images integrated for less than 1 minute. 
    (b) FWHM of the PSF for long PL images (orange) and the Gaussian convolution observed in magnetic images (purple) of Fe\textsubscript{2}O\textsubscript{3} MNPs on diamond - both integrated for over 3 hours. 
    Also shown in (a,b) are plots of the FWHM of the theoretical diffraction limit (black, solid), and calculated FWHM of the PSF through a 75 µm thick diamond (gray, dashed).
    (c) Heatmaps of the FWHM of the central peak of the PSF simulated using the Gibson-Lanni model and (d) collection efficiency as a function of the diamond thickness and NA of the objective for emission wavelength of 650 nm.
    The solid white and black lines in (d) denote $S$ = 0.45 and 0.8 and signifies where the FWHM and the collection efficiency saturate, respectively, and represents an optimal region.}
    \label{fig:FWHM_vs_NA}
\end{figure}

To evaluate the optical and magnetic resolution of our system, without the impact of standoff, we imaged 50\,nm Fe\textsubscript{2}O\textsubscript{3} magnetic nanoparticles (MNPs) deposited directly onto our diamond sensor (Fig.~\ref{fig:Det_res}(a)) with each objective. The analysis for the 20x objective is shown in Fig.~\ref{fig:Det_res}, as an example. The MNPs appear as bright blob-like features in the images of photoluminescence (PL) from the NV layer (Fig.~\ref{fig:Det_res}(b)) and were fitted using a 2D Gaussian function to determine the lateral PSF at focus~\cite{Cole2011} (Fig.~\ref{fig:Det_res}(c)). The optical resolution of each objective was taken as the smallest reproduced full width at half maximum (FWHM), average of $x$ and $y$, that provided a good fit of the data, e.g. $\approx0.82$\,µm for the 20x objective. A similar analysis was performed for 140\,nm fluorescent nanodiamonds (FNDs) on a coverslip (Fig.~\ref{fig:Det_res}(d)) to estimate the optical resolution of our system in the absence of the diamond slab ($\approx0.79$\,µm with the 20x objective), which will be used for comparison in Fig.~\ref{fig:FWHM_vs_NA}. 

To acquire a magnetic image, the PL of a region of the diamond was imaged as the MW frequency was swept to produce an ODMR spectrum at each pixel in the camera. The sweep was repeated and the images at each frequency integrated for a few hours to improve the signal-to-noise. A bias magnetic field was applied along one of the four NV axes in the diamond ($B_{\rm NV}^0=7.88$\,mT) — to accelerate the rate of measurement, only the corresponding resonances were measured. The peaks were fitted with Lorentzian functions to accurately determine the frequency of the resonances, $f_{1}$ and $f_{2}$ and used to calculate the magnetic field projected onto the NV axis $B_{\rm NV}=\frac{f_2-f_1}{2\gamma_e}$, where $\gamma_e=28$\,MHz/mT is the electron gyromagnetic ratio. To attain only the stray field, the bias magnetic field was removed $\Delta B_{\rm NV}=B_{\rm NV}-B_{\rm NV}^0$ as shown in Fig.~\ref{fig:Det_res}(e-g). 

To quantify the magnetic resolution we begin by calculating the stray field from the MNPs at a specific standoff distance between the particles and the diamond surface plus a range of depths for the NVs, weighted by the relative concentration at each depth. This corresponding field is broadened with a 2D Gaussian to account for the optical PSF. Line profiles across the main lobes in the experimental NV projected stray field images were fitted using corresponding line profiles from the simulated images. The FWHM of the Gaussian broadening was used as a fitting parameter (0.80 µm in Fig.~\ref{fig:Det_res}(g)) and defines the impact of the optical resolution on the resolution in magnetic images. Note, the optical broadening here may be larger than that obtained in a simple PL image due to the much longer measurement duration affected by drift, thermal expansion and mechanical instability, as commonly observed in microscopes~\cite{Acher2023}.

A summary of our spatial resolution analysis as a function of the NA of the objective is provided in Fig.~\ref{fig:FWHM_vs_NA}(a-b). We define the resolution using the FWHM of our 2D Gaussian fits. We split our analysis into short and long duration to isolate the intrinsic limitations from the additional loss in resolution associated with longer durations. First, we demonstrate that our QDM is diffraction limited by determining the PSF from PL images of the FNDs (Fig.~\ref{fig:Det_res}(d)) for each objective - we observe a strong agreement with the theoretical PSF defined by the Airy disk where $\text{FWHM} = 0.51\,\text{NA}/\lambda$. To quantify the PSF in the non-design case (i.e. when the diamond slab is present in the imaging path), we analyse the PL images from MNPs on diamond as exemplified in Fig.~\ref{fig:Det_res}(c). For short measurement durations (Fig.~\ref{fig:FWHM_vs_NA}(a)), there is excellent agreement with the PSF from the FND (within 5\%) for all except the high NA = 1.3 objective for which the PSF is 54\% larger, which we will address later. Since ODMR analysis may require hours of measurement to achieve a good signal-to-noise ratio in the magnetic images we repeated the PSF analysis using the integrated PL images from the ODMR measurements. The PSF increased compared to the short-term PL by 7\% on average with the deviation generally larger for high NA objectives. Likewise, the FWHM of the Gaussian convolution used to estimate the optical resolution component of the magnetic resolution shows good agreement with the long timescale PSF and within 10\% of the diffraction limit for NA $\leq$ 0.6 but deviates at higher NA. The plateau in the FWHM to 0.6\,µm at NA $\geq$ 0.6 indicates the resolution in this regime is limited by mechanical instabilities on this scale~\cite{Acher2023}. Although not performed here, this can be readily corrected by saving the ODMR images at regular intervals and post-processing to realign the images using standard drift correction procedures, at the cost of measurement speed.

We now consider how imaging through diamond may deteriorate the optical resolution using the model developed by Gibson and Lanni~\cite{Gibson1992}. Under design conditions, the object plane is immediately below the coverslip and is in focus at the detector. For a point source at some depth from the coverslip the specimen stage is shifted along the optical axis towards the lens until the object plane is in focus at the fixed detector. This reduces the immersion layer thickness and alters the optical path from the design case. Differences in the wavefield at the back focal plane of the objective between the design and actual case may cause additional aberration that degrades the resolution and contrast. To quantify the impact of this aberration we use the code written by Anthony et al.~\cite{Anthony2019} that employs the Gibson-Lanni model, based on scalar diffraction, to explore the influence of objective NA and diamond thickness. The PSF was calculated for both air and oil immersion objectives (refractive index $n_i = 1.52$) and two useful parameters were determined - the FWHM of the central peak and the Strehl ratio $S$. The latter is a figure of merit of the quality of images formed in optical systems that scales between 0 and 1 with values $>\,0.8$ generally being considered diffraction limited~\cite{Nishimura2024}. It is defined as the ratio of the peak intensity in the aberrated PSF and the maximum attainable intensity in a diffraction limited system. Heatmaps of the FWHM and Strehl ratio are given in Fig.~\ref{fig:FWHM_vs_NA}(c) and Fig. S3, respectively. For no diamond present (diamond thickness = 0\,µm) and low NA for finite diamond thicknesses, the PSF matches the design case (diffraction limited). However, for finite thicknesses of diamond, as the NA is increased the central peak broadens and adjacent rings become more prominent and the FWHM saturates (corresponding to $S \approx 0.45$), although the Strehl ratio continues to drop. For the case of a 75 µm thick diamond, we have plotted the variation in the FWHM with respect to NA, Fig.~\ref{fig:FWHM_vs_NA}(a-b). For NA = 0.8 and 1.3 the model predicts FWHM = 0.47 and 0.35\,µm, respectively, which is in reasonable agreement with the experimentally observed FWHM = 0.47 and 0.43\,µm for the short duration measurement of the MNPs on diamond. 

Another important quantity that we can ascertain from this model is the collection efficiency $CE$. While it is not easily determined experimentally, efficient readout of the NV fluorescence is critical for magnetic field sensing with the magnetic sensitivity of a QDM scaling as $\eta \propto \frac{1}{\sqrt{CE}}$~\cite{Barry2020}. The collection efficiency depends on competing processes including the acceptance angle of the objective, refraction at the various interfaces and loss in intensity due to aberrations, as defined by the Strehl ratio. The fraction of emitted light that is collected by an objective is given by $F = \frac{1}{2}(1-\sqrt{(1-(\text{NA}/n_{d})^2)})$ where $n_{d}$ is the refractive index of diamond~\cite{Prawer2014}. The angle dependent reflection losses at the diamond|glass and glass|immersion interfaces $R$ can be estimated using the Fresnel equations. Combining these together we arrive at $CE = S\times F\times (1 - R)$ which is plotted for air and oil objectives against NA and diamond thickness in Fig.~\ref{fig:FWHM_vs_NA}(d). At low NA or very low diamond thicknesses, the Strehl ratio is close to unity and the collection efficiency is limited by the NA. For finite diamond thicknesses, increasing the NA leads to an increase in the collection efficiency until $S\,=\,0.8$ (black line, Fig.~\ref{fig:FWHM_vs_NA}(d)) after which $CE$ drops significantly and is limited by the Strehl ratio due to aberrations. The band between $S$\,=\,0.45 (white line) and $S\,=\,0.8$ provides a useful target with an excellent compromise between collection efficiency and spatial resolution. On average, the objectives with $S$\,=\,0.45 exhibit 12\% lower FWHM but 25\% lower collection efficiency than $S = 0.8$. For a diamond thickness of 50\,µm, an air objective with NA\,=\,0.75 is a good compromise giving FWHM\,=\,0.45\,µm and a $CE=1.2\times10^{-2}$. Increasing the thickness to 300\,µm, NA\,=\,0.5 (air) now provides a good compromise resulting in FWHM\,=\,0.68\,µm and a $CE=5.9\times10^{-3}$, while using an oil objective (NA\,=\,1.3) would reduce FWHM to 0.47\,µm but reduce $CE=1.8\times10^{-3}$. This modelling highlights the importance of matching the objective with the diamond chip in order to achieve the best resolution and/or measured fluorescence intensity possible. The predicted performance of the objectives relevant to this work are provided in Fig. S4.

A quick remark on the validity of the PSF study presented here. We opted to use the Gibson-Lanni method because it is computationally inexpensive and readily applicable albeit with a limited accuracy. A more rigorous study into the loss of optical resolution when measuring through diamond, including the effect of off-centre imaging, can be found elsewhere~\cite{Nishimura2024}. However, for diamond thickness =\,400\,µm and NA\,$>$\,0.8, we compared the PSF with those calculated using the vectorial diffraction approach developed by Aguet et al.~\cite{Aguet2009} and found that on average the Gibson-Lanni method underestimated the FWHM and $S$ by only 10\% and 13\% for the oil objective ($\le\,5\%$ for both FWHM and $S$ for air objectives). Therefore, our approach is valid and justified for on-axis analysis. 

\begin{figure}[tb!]
    \includegraphics[width=8.5cm]{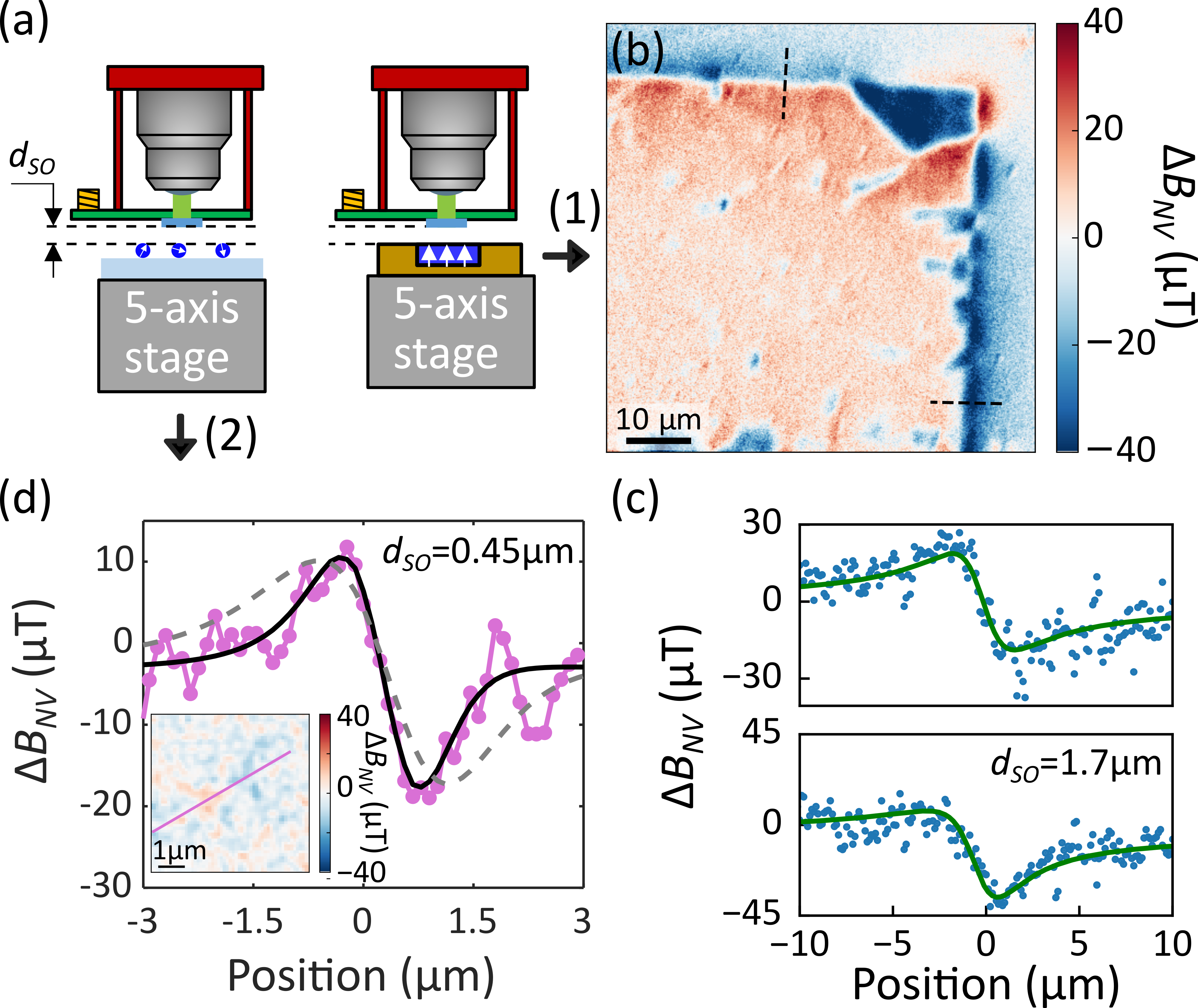}
    \caption{\textbf{Determination of the standoff distance between the diamond sensor and specimen.}
    (a) Schematic of the QDM measuring Fe\textsubscript{2}O\textsubscript{3} MNPs on glass (left) and a magnetic thin film of CoFeB (1 nm thick) on a piece of silicon (right). 
    (b) Magnetic field image of the corner of a CoFeB pad. Dashed lines indicate where line profiles were taken.
    (c) Example line profiles and fits across the edges of the CoFeB film, with a standoff distance of 1.7\,µm. All PL and magnetic images were taken with the 50x objective (NA\,=\,0.8).
    (d) Example line profile of a Fe\textsubscript{2}O\textsubscript{3} MNP and corresponding simulated magnetic field with $d_{SO}\,=\,0.45$\,µm (black, solid line) and 1.0\,µm (gray, dashed line). Inset: Magnetic field image of the MNP.}
    \label{fig:dso}
\end{figure}

Having considered the optical contribution, we now turn our attention to the impact of standoff distance on the lateral resolution. To quantify the standoff distance $d_{SO}$ that we can achieve using our new sensor holder, we imaged the stray field from two different samples. (1) An ultrathin (1 nm) magnetic pad of CoFeB that has been stored in a standard laboratory for years and (2) Fe\textsubscript{2}O\textsubscript{3} MNPs freshly deposited onto a clean coverslip, as depicted in Fig.~\ref{fig:dso}(a). The former is likely to have debris on the surface that limits standoff while the latter is expected to be significantly cleaner. The sample was approached while observing interference fringes that appeared between the diamond and sample surfaces. The fringes expand as the gap between surfaces reduces. The tilt of the sample was fine-tuned until only a single band was observed in the field of view to minimise $d_{SO}$, following the method described by Abrahams et al.~\cite{Abrahams2021}. 

To analyse the image of the magnetic pad (Fig.~\ref{fig:dso}(b)), least-square fits were performed for a collection of line cuts across two perpendicular edges with the theoretical stray field calculated at a specific standoff and convolved with a Gaussian function (FWHM = 0.51\,µm, to emulate the optical broadening from the 50x NA = 0.8 objective used here), as described elsewhere~\cite{Abrahams2021}. From 150 sets of line profiles (examples in Fig.~\ref{fig:dso}(c)), we estimate a mean standoff $\langle d_{SO} \rangle\,=\,1.9$\textpm 0.1\,µm, in good agreement with our previous results. In the case of the MNPs on the coverslip, we performed a similar analysis as described earlier for determining the magnetic resolution of MNPs, except with the $d_{SO}$ used as a fitting parameter and the Gaussian broadening fixed (FWHM\,=\,0.51\,µm). We observe that $d_{SO}$ reduces to 0.45\,µm, as shown in Fig.~\ref{fig:dso}(d), with the addition of a weighted average thickness of 50\,µm for the NV layer, results in features in the magnetic image with a FWHM\,=\,1.2\,µm, a significant improvement on the 4\,µm we achieved on an earlier QDM design~\cite{Abrahams2021}.

The best standoff distance reported here of 0.45\,µm across a 1.5\,mm\,×\,1.5\,mm interface demonstrates the effectiveness of our QDM design. For this standoff distance, a 50x NA\,=\,0.8 air objective is a good choice for a 75\,µm thick diamond to achieve optical broadening of a similar order (0.51\,µm) while maintaining good collection efficiency. In order to further improve both the spatial resolution and collection efficiency it is necessary to use much thinner diamonds, other groups~\cite{Riedel2020-kd, Guo2021} have employed diamond membranes with thicknesses $\approx1$\,µm. In this case, oil objectives would lead to a PSF FWHM\,=\,0.26\,µm for NA\,=\,1.3 and even 0.2\,µm for NA\,=\,1.7. Using the latter objective, magnetic imaging with an overall spatial resolution of $<\,500$\,nm could be achieved with standoff distance of 200\,nm, which is feasible with cleaner and/or flatter interfaces. For less clean or rougher interfaces limitting standoff distance $\approx2$\,µm, as in Fig.~\ref{fig:dso}(b, c), there would be only a slight loss in overall resolution with a thicker diamond and lower NA. For example, a 400\,µm thick diamond measured with an NA\,=\,0.45 objective gives a PSF FWHM\,=\,0.74\,µm with almost five times greater collection efficiency ($CE = 5.4\times10^{-3}$ compared to $1.2\times10^{-3}$ for NA\,=\,0.8). A final consideration for improved imaging, particularly at high resolution, is drift/instability of the instrument throughout long measurements, as evidenced by the difference between the FWHM of the PSF for the short and long duration analysis in Fig.~\ref{fig:FWHM_vs_NA}. Implementing an auto-focus and automated drift correction could significantly narrow that gap.

In summary, we have developed a co-axial sensor holder that allows for simple and repeatable interfacing between the diamond sensor and sample while being fully compatible with high NA optics. We demonstrate our best spatial resolution of 1.2\,µm with an NA\,=\,0.8 air objective, resulting from a standoff distance of 0.45\,µm and optical broadening of 0.51\,µm. Our analysis on the dependence of the optical resolution and collection efficiency on objective and diamond thickness allows researchers to match them both for optimal imaging for their application, and will guide future efforts to further improve the spatial resolution. The capability to routinely achieve high spatial resolutions across wide fields of view will improve the utility of the technique and promote the adoption of widefield quantum diamond microscopy for a broad range of applications.

\section*{Supplementary Material}
See supplementary material for additional details of the diamond sensors, simulations of the magnetic field from a MNP, PSF analysis and additional photographs and schematics of the setup.

\begin{acknowledgments}
This work was supported by the Australian Research Council (ARC) through grants FT200100073, DP220100178, and DE230100192. 
I.O.R. is supported by an Australian Government Research Training Program Scholarship.
S.C.S gratefully acknowledges the support of an Ernst and Grace Matthaei scholarship.
\end{acknowledgments}

\section*{Author Declarations}

The authors have no conflicts to disclose.

\bibliography{refs}

\clearpage

\onecolumngrid

\begin{center}
\textbf{\large Supplementary Material}
\end{center}

\setcounter{equation}{0}
\setcounter{section}{0}
\setcounter{figure}{0}
\setcounter{table}{0}
\setcounter{page}{1}
\makeatletter
\renewcommand{\theequation}{S\arabic{equation}}
\renewcommand{\thetable}{S\arabic{table}}
\renewcommand{\thefigure}{S\arabic{figure}}

\section{Diamond details} \label{SI:diamond}

Two diamonds were used throughout this work. For all but the standoff distance study, the measurements were performed using the following diamond. A 2\,mm × 1\,mm diamond sensor was cut and prepared from a 4\,mm × 4\,mm × 75\,µm type-Ib single crystal substrate grown using high-pressure, high-temperature (HPHT) with {100}-oriented polished faces, purchased from DDK. To form a dense NV ensemble layer the diamond was implanted with 230 keV $^{32}\rm S$ ions at a dose of $3.7\times10^{11}\,\rm{ions/cm^{2}}$. We performed full cascade Stopping and Range of Ions of Matter (SRIM) Monte Carlo simulation to estimate the depth distribution of the created vacancies with depths varying from 0 to 200\,nm, peaking at 100\,ppm vacancies at $\approx 120$\,nm (assuming no dynamic annealing effects). 

The second diamond sensor, used for the standoff distance study in Fig.\ 4, is 1.5\,mm × 1.5\,mm cut from 4.5\,mm × 4.5\,mm × 50\,µm diamond (from DDK) that was implanted with 400 keV $\rm Sb$ at a dose of $2.0\times10^{11}\,\rm{ions/cm^{2}}$ to form NV at depths varying from 0 to 150 nm with a peak of about 150\,ppm vacancies at $\approx 65$ nm and weighted average $\approx 50$\,nm. 

\section{Experimental setup} \label{SI:exp_setup} 
The experimental setup is shown in Fig. 1 of the main text. Two different light sources were used, an LED and a laser. The LED provides uniform illumination for widefield imaging to identify regions of interest and for analysis of the optical spatial resolution of the system. The higher flux of the laser results in improved contrast in ODMR spectra and higher magnetic sensitivity. It was used for magnetic imaging with high NA objectives (NA\,$>$\,0.5) in Fig. 3 and Fig. 4 in the main text. A continuous-wave (CW) laser excitation was produced by a $\lambda=532$\,nm solid-state laser (Laser Quantum Opus 2 W). The laser beam was expanded (3x) and focused using a $f = 150$\,mm lens to the back aperture of the objective lens via a dichroic mirror. The other light source, an LED (Thorlabs: M530L4) was predominately used in an epi-fluorescent fashion as shown in Fig. 1. In this case, a $f = 20$\,mm aspheric lens collimated the light, filtered using a short-pass filter which was focused onto the back of the objective using a $f = 75$\,mm achromatic lens. To improve the photoluminescence (PL) contrast of nanoparticles, particularly for low NA objectives, the LED light source was set up in transmission arrangement. In this case, the LED was mounted co-axial, below and pointed towards the objective. The light was collimated using a $f = 16$\,mm lens, filtered with a short-pass filter and focused at the sample using $f = 16$\,mm lens. The PL from the NV layer was collected using an objective, separated from the excitation light with a dichroic mirror, filtered using a 650\,nm long-pass filter, and imaged using a tube lens ($f = 200$\,mm) onto a CMOS camera (Basler acA2040-90um USB3 Mono). We used a collection of objectives as summarised in Table \ref{tab:table1}. The Olympus objectives are designed for a $f = 180$\,mm  tube lens so experience a slight magnification in our setup relative to their design magnification.  

Microwave excitation was provided by a signal generator (Windfreak SynthNV Pro) gated using an IQ modulator (Texas Instruments TRF37T05EVM) and amplified (Mini-Circuits ZHL-16W-43+). The output of the amplifier is directly connected to the PCB holding the diamond using a coaxial cable with a MMCX connector. A pulse pattern generator (SpinCore PulseBlasterESR-PRO 500 MHz) was used to gate the microwave (MW) and to synchronise the image acquisition. The optically detected magnetic resonance (ODMR) spectra of the NV layer were obtained by sweeping the MW frequency, taking a reference image (MW off) for each frequency in order to produce a normalized spectrum removing common-mode PL fluctuations. The exposure time for each camera frame was varied from 20 to 500\,ms depending on the objective used (to ensure all measurements are photon shot noise limited rather than camera readout noise limited), and the entire sweep was repeated thousands of times to improve the signal to noise ratio. A bias magnetic field of a few mT was applied using a permanent magnet, roughly aligned with one of the four possible NV orientation classes (corresponding to the $\langle 111 \rangle$ crystal directions). We swept the MW frequency in order to interrogate this mostly aligned NV family only. All measurements were performed in an ambient environment at room temperature.

\begin{table*}[tb!]
\caption{\label{tab:table1} Details of the objectives. For the Olympus objectives, the actual magnification with the tube lens used in our setup is indicated in brackets. Here NA is the numerical aperture, $n_i$ is the refractive index of the immersion medium, and W. D. is the working distance.}
\begin{ruledtabular}
\begin{tabular}{lccccc}
 & Magnification & Immersion & NA & $n_i$ & W. D. (mm) \\
\hline
Olympus UPLFLN 4x & 4 (4.44) & Air & 0.13 & 1 & 17\\
Olympus UPLFLN 10x & 10 (11.1) & Air & 0.3 & 1 & 10.6\\
Nikon CFI S Plan Fluor ELWD 20xC & 20 & Air & 0.45 & 1 & 6.9-8.2\\
Nikon CFI S Plan Fluor ELWD 40xC & 40 & Air & 0.6 & 1 & 2.8-3.6\\
Nikon CFI TU Plan Apo Epi 50x & 50 & Air & 0.8 & 1 & 2\\
Nikon CFI Super Fluor 40x Oil & 40 & Oil & 1.3 & 1.52 & 0.19 \\
\end{tabular}
\end{ruledtabular}
\end{table*}

\section{Simulated magnetic field imaging and fitting} \label{SI:sim_magnetic}
Simulations of the experimentally measured stray field from the Fe\textsubscript{2}O\textsubscript{3} magnetic nanoparticles (MNPs) were performed using the method develop by Tetienne et al.~\cite{tetienneProximityinducedArtefactsMagnetic2018} This involves simulating the stay field from the nanoparticles, projecting the field onto NV axes, simulating the corresponding ODMR spectrum, 2D Gaussian broadening each single-frequency image in the simulated ODMR to account for the optical PSF, and fitting the resulting ODMR spectrum at each pixel to generate a magnetic image. We begin by considering a cube-shaped nanoparticle of 50\,nm\textsuperscript{3} with saturation magnetisation $M_S \approx 10^6$\,A/m. Each particle is assumed to have a single magnetic domain and the direction of the magnetisation vector was adjusted to replicate the magnetic image of a specific particle. Four NV axes were set in the simulations to match the NV axes in the diamond sensor and in both simulations and experiments a bias field of $\approx 7.5$\,mT aligned to one of the axes was applied. The stray field was calculated at a specific standoff distance from the NV layer, the magnetic field projection onto the NV axis and the corresponding ODMR spectrum maps were calculated. Since the NV layer is about 200\,nm with a depth varying concentration, this was repeated in steps of 10\,nm to provide the relative concentration weighted ODMR spectrum at each pixel. 

The experimental magnetic field images from specific particles were simulated using the above method and fitted, with the magnetisation direction and saturation and the full width at half maximum (FWHM) of the Gaussian convolution FWHM\textsubscript{mag} used as fitting parameters. Simulating a magnetic field image of a single particle took several minutes so fitting a 2D magnetic experimental image with simulated magnetic image was too computationally demanding and would need to be repeated for multiple particles, for each of the objectives used here. Instead we opted to least-squares fit a line profile across the main lobes in experimental image $\Delta B_{NV}^{exp}$ with a corresponding line profile across a simulated magnetic $\Delta B_{NV}^{sim} = f(x)$ using linear transforms of the simulated line profiles as initial fitting parameters, where $x$ is the position along the line profile. The fitting function involves two steps: a new x-scale is created by dilating and offsetting the original x-scale, $x' = p_1 x + p_2$, and $y_{sim}$ is interpolated on this new x-scale with dilation and offset applied, $\Delta B_{NV}^{fit} = p_3 f(p_1x+p_2)+p_4$, where $p$ indicates fitting parameters. Suitable offsets are applied to the simulated line profiles based on the parameters $p_2$ and $p_4$ and the least-square fit is repeated for the same simulated image, often multiple times until $p_2$ and $p_4$ $\approx 0$. The dilation parameters $p_1$ and $p_3$ are used to inform on necessary changes to the FWHM\textsubscript{mag} and $M_S$ in the simulations; these parameters are updated and the simulations rerun. This process is repeated until $p_1$ and $p_3$ $\approx 1$ which usually requires only a few iterations. The method allows us to determine the FWHM\textsubscript{mag} to a precision of 0.01 µm.  

\begin{figure}[tb!]
    \includegraphics[width=10.5cm]{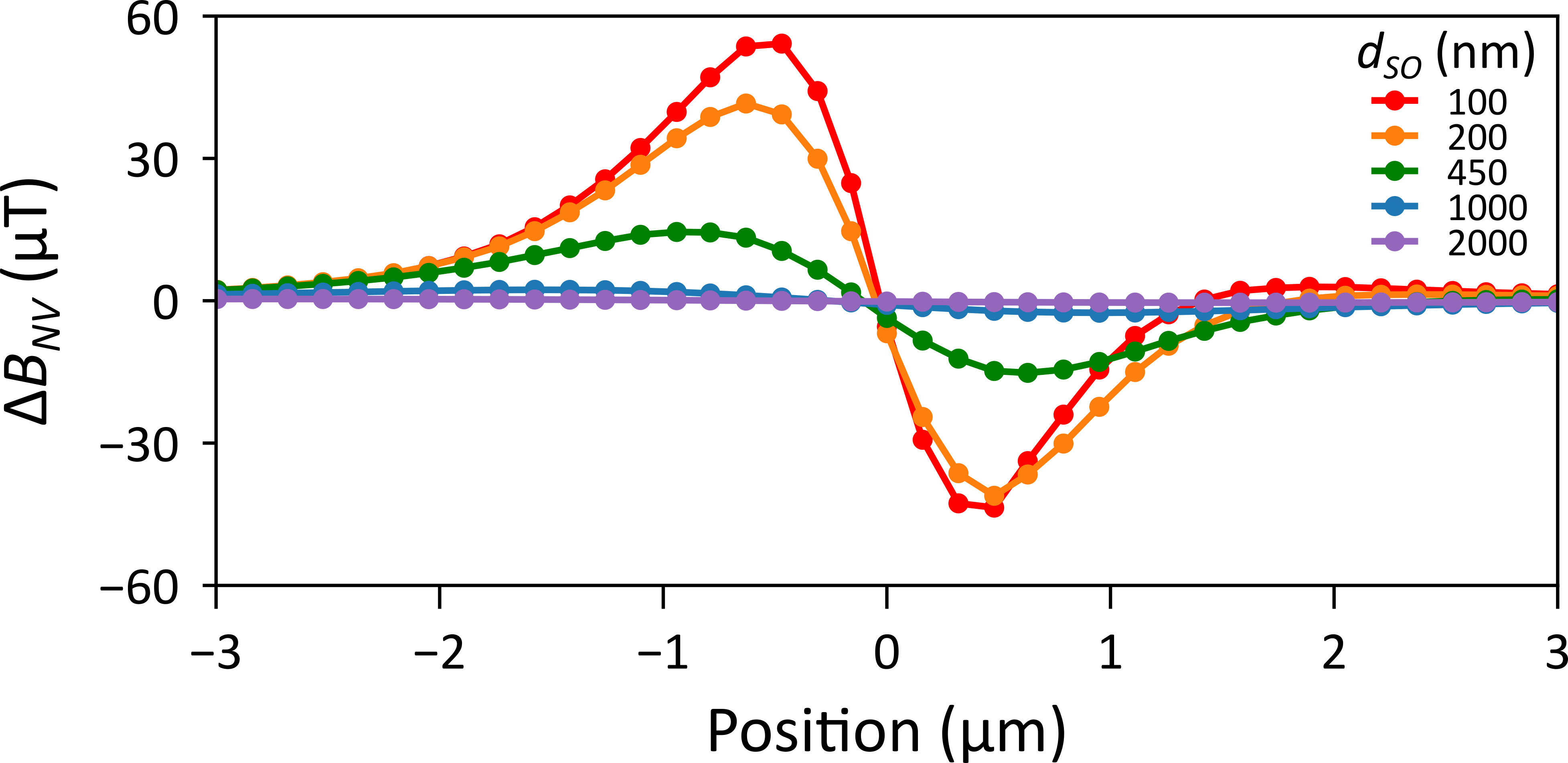}
    \caption{\textbf{Simulated magnetic stray field line profiles of a Fe\textsubscript{2}O\textsubscript{3} MNP.} Similar to the simulation presented in Fig. 4(d) with $d_{SO} =$ 0.1, 0.2, 0.45, 1 and 2\,µm, assuming optical broadening FWHM = 0.51\,µm  $M_S\,=\,4\times 10^{5}$\,A/m.}
    \label{fig:BNVvsdso}
\end{figure}

In order to demonstrate the influence of the standoff distance on magnetic field strength and spatial resolution simulations of the magnetic stray field from an Fe\textsubscript{2}O\textsubscript{3} MNP shown in Fig. 4 in the main text has been repeated with standoff distances varying from 0.1 to 2\,µm are shown in Fig.~\ref{fig:BNVvsdso}. The peak-to-peak stray field reduces from 98 to 0.8\,µT, while the FWHM of features increases from 0.6 to 4\,µm. Reducing the standoff is important for improving the spatial resolution and accuracy of measurements of the magnetic field.

\section{Sensor holder designs} \label{SI:holder}

\begin{figure}[tb!]
    \includegraphics[width=10.5cm]{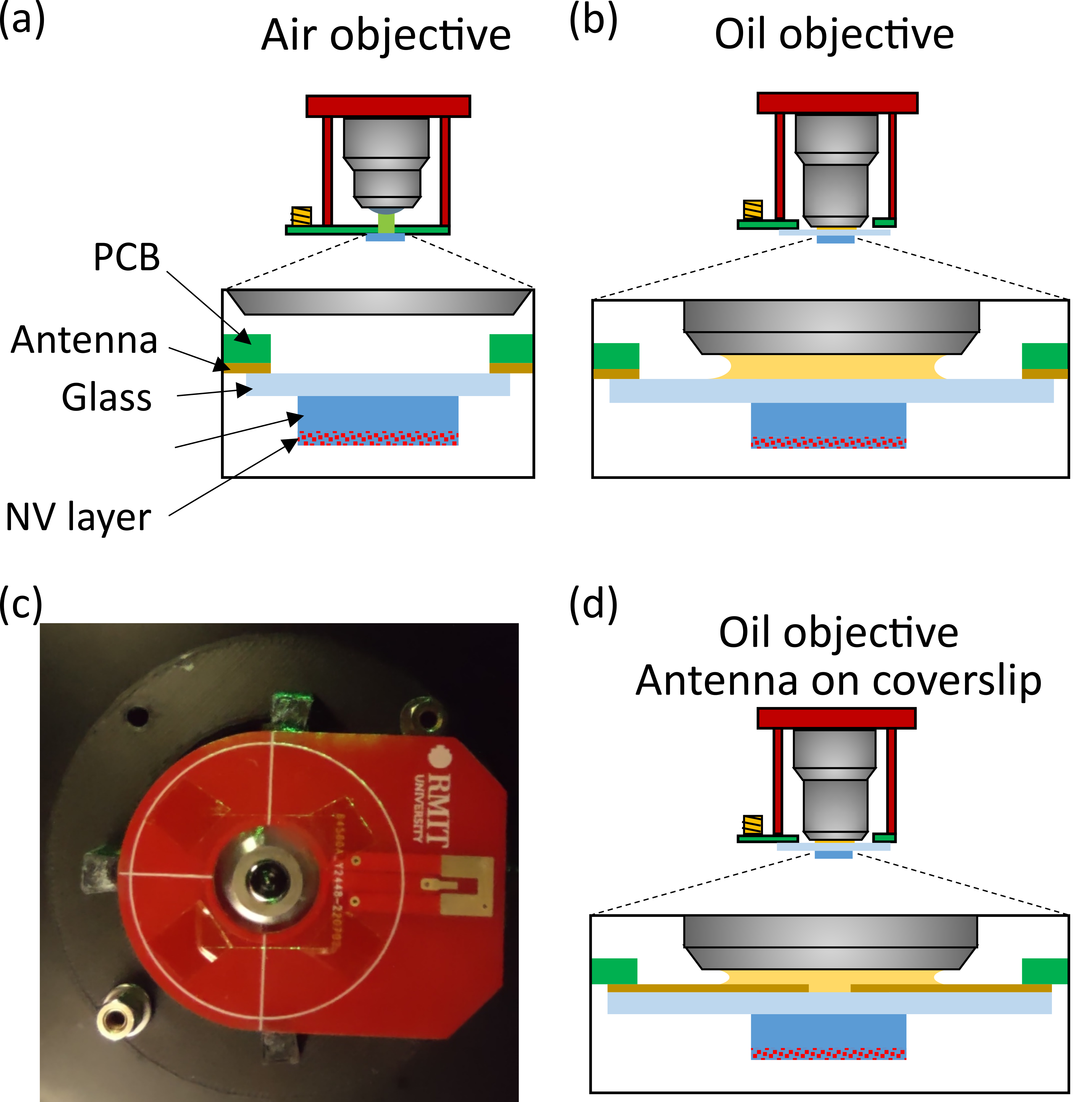}
    \caption{\textbf{Schematics and photograph of our widefield probe.} (a) Schematics of the widefield probe including a PCB with a 3.2\,mm integrated loop antenna in the PCB for air objectives and (b) 12\,mm loop for oil objectives. (c) Photograph of our realisation of (b) with a 12 mm loop around the 40x oil objective. (d) Schematic of an alternate design in which the MW resonator is deposited onto the coverslip with contacted to the PCB.}
    \label{fig:Schem_probe}
\end{figure}

The heart of our quantum diamond microscope design is the diamond sensor holder which consists of a diamond chip glued to a coverslip attached to a PCB, fixed to a holder co-axial to the optics. The PCB has an integrated MW loop antenna co-axial with an aperture. For air objectives we use a loop with a 3.2\,mm diameter, Fig.~\ref{fig:Schem_probe}(a). However, for oil objectives, the thickness of the PCB (0.8\,mm) prevents most oil objectives, due to small working distances ($\approx 0.2$\,mm), from achieving focus at the NV layer in the diamond. In this case, the aperture is increased to permit the tip of the objective with some space to position/tilt the diamond; we use a 12\,mm loop (Fig.~\ref{fig:Schem_probe}(b-c)). Unfortunately, for the same MW power, the flux of MW at the diamond is significantly reduced which may affect sensitivity during magnetic imaging. An alternate design, to allow smaller loops, is to deposit a gold resonator directly onto the coverslip and use solder and/or conductive paste to make contact to a PCB with suitable RF connectors (Fig.~\ref{fig:Schem_probe}(d)). To prevent damage to the resonator it is preferably faced towards the objective.

\section{Simulated point-spread function analysis} \label{SI:PSF}
\begin{figure}[tb!]
    \includegraphics[width=6.5cm]{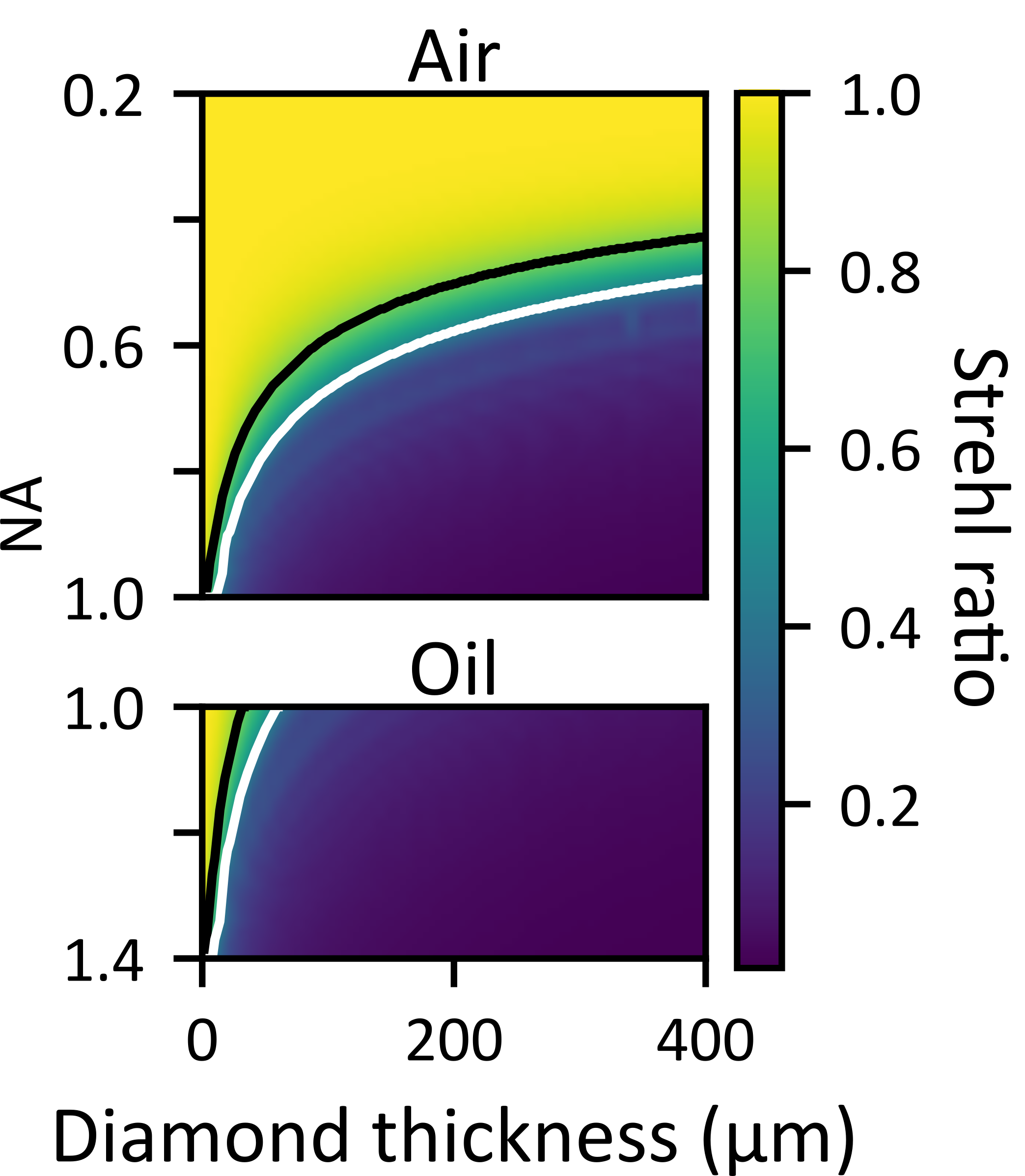}
    \caption{\textbf{Point-spread function analysis for air and oil objectives.} Heatmap of the Strehl ratio of the PSF simulated using the Gibson-Lanni model as a function of the diamond thickness and NA of the objective for emission wavelength of 650\,nm. The solid white and black and line in denotes \textit{S}\,=\,0.45 and 0.8, respectively.}
    \label{fig:S_heatmap}
\end{figure}

Point spread functions were calculated using the code written by Anthony et al.~\cite{Anthony2019} based on the Gibson-Lanni model for refractive index mismatch induced aberration. We assume a fixed glass coverslip thickness of 0.17\,mm with refractive index $n_{i} = 1.52$ but allowed the immersion thickness to vary. The refractive index of diamond $n_{d} = 2.41$. To identify the optical focus, the calculation is performed for multiple immersion thicknesses and focus is determined as the greatest Strehl ratio. Optimising the Strehl ratio led to a modulation in the FWHM with diamond thickness where $S < 0.4$. To minimise this, the immersion thickness was prevented from reducing with increasing thickness, providing a compromise between focusing using the Strehl ratio and FWHM. A heatmap of the strehl ratio as a function of NA and diamond thickness is provided in Fig. \ref{fig:S_heatmap}. The calculation was repeated at higher spatial resolution and the FWHM is determined from the position where the intensity of the peak is half the maximum. This was faster to calculate and provided similar results to least-squares fitting the simulated PSF. 

The PSF calculated using the code by Anthony et al.~\cite{Anthony2019} were validated using more the rigorous vectorial diffraction approach and code developed by Aguet et al.~\cite{Aguet2009}, for diamond thickness = 400\,µm and NA $>$ 0.8. Similar refractive indices and coverslip thickness were used in this calculations. The maximum Strehl ratio was used to identify focus and the FWHM was determined by considering half the peak height.  

\begin{figure}[tb!]
    \includegraphics[width=10.5cm]{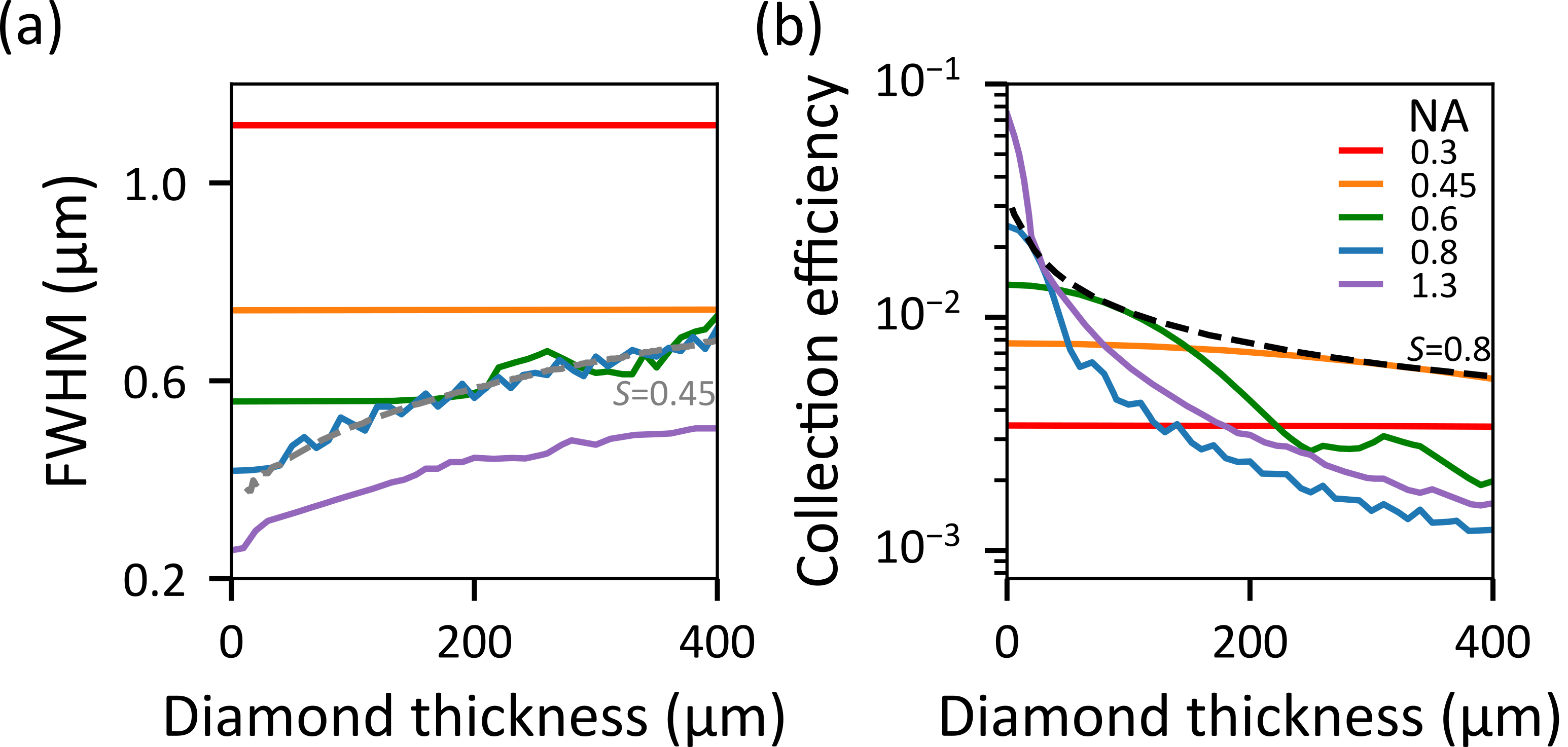}
    \caption{\textbf{Performance of various objectives imaging through diamond.} (a) FWHM of the central peak of the PSF simulated using the Gibson-Lanni model and (b) collection efficiency (defined in the main text) as a function of diamond thickness for various objective NA. The dashed grey and black lines in (a) and (b) denotes $S = 0.45$ and 0.8 and correspond to roughly where the FWHM and collection efficiency saturate, regardless of NA for air objectives.}
    \label{fig:FWHM_vs_t}
\end{figure}

The predicted performance of objective used to collect experimental data, with NA $>$ 0.2, is summarised in Fig.~\ref{fig:FWHM_vs_t}. At low thicknesses, the FWHM reduces with NA. For each NA, the FWHM remains constant with diamond thickness until $S \approx 0.45$ (indicated by the grey dashed line) at which point it scales sublinearly with the thickness, albeit with some irregular modulation. A similar pattern can be observed in the collection efficiency, for each NA the collection efficiency shows only a slight reduction with thickness, until S approaches 0.8 and the collection efficiency drops more sharply. At low thicknesses, the collection efficiency improves with NA. The plots in Fig.~\ref{fig:FWHM_vs_t} highlight the importance of pairing the objective - simply choosing the highest NA may lead to sub-optimal imaging. 

\end{document}